\documentclass[aps,prl,twocolumn,showpacs,superscriptaddress,groupedaddress]{revtex4}
\usepackage[centertags]{amsmath}
\usepackage{amsmath}
\usepackage[dutch,british]{babel}
\usepackage{amsfonts}
\usepackage{amssymb}
\usepackage{amsthm}
\usepackage{newlfont}
\usepackage{color}
\usepackage{subfig}
 \usepackage{bbm}

\usepackage{stmaryrd}
\usepackage{mathrsfs}
\usepackage{fancyhdr}
\usepackage{graphicx}
\usepackage{fancybox}
 \usepackage{psfrag}
 
 \usepackage{multirow}

\newcommand{\e}{\mathrm{e}}
\renewcommand{\d}{\mathrm{d}}
\newcommand{\str}{ |}
\newcommand{\norm}{ ||}
\newcommand{\bbZ}{\mathbb{Z}}
\theoremstyle{plain}
\newtheorem{theorem}{Theorem}

\newcommand{\ad}{\mathrm{ad}}

\begin{document} 

\title{Adiabatic theorem for 
quantum spin systems}
\author{S.~Bachmann}
\affiliation{
Mathematisches Institut der Universit{\"a}t M{\"u}nchen, Germany
}
\author{W.~De Roeck}
\affiliation{
Instituut voor Theoretische Fysica, KU Leuven, Belgium
}
\author{M.~Fraas}
\affiliation{
Instituut voor Theoretische Fysica, KU Leuven, Belgium
}

 \pacs{63.10.+a, 05.30.-d}

\date{\today}
\begin{abstract}
The first proof of the quantum adiabatic theorem was given as early as 1928. Today, this theorem is increasingly applied in a many-body context, e.g.\ in quantum annealing and in studies of topological properties of matter. In this setup, the rate of variation  $\varepsilon$ of local terms is indeed small compared to the gap, but the rate of variation of the total, extensive Hamiltonian, is not.   Therefore, applications to many-body systems  are not  covered by the proofs and arguments in the literature.  In this letter, we prove a version of the  adiabatic theorem for gapped ground states of interacting quantum spin systems, under assumptions that remain valid in the thermodynamic limit. As an application, we give a mathematical proof of Kubo's linear response formula for a broad class of gapped interacting systems. We predict that the density of non-adiabatic excitations is exponentially small in the driving rate and the scaling of the exponent depends on the dimension.
\end{abstract}

\maketitle

\subsection{Introduction}
The premise of adiabatic theory that slow driving keeps a system close to its ground state is central to our understanding of gapped quantum systems.
It underlies the discussion of the quantum Hall effect (QHE), non-dissipative response theory or the classification of gapped phases of matter. It is also used to justify adiabatic control methods employed in manipulation of artificial states of matter.  Despite its importance no proof of the adiabatic theorem for many-body (extended) systems is known. Phrased differently, there is no version of adiabatic perturbation theory that respects locality.

In this letter we fill this gap by deriving an adiabatic theorem for interacting, extended systems that remain gapped in the thermodynamic limit. This covers some of the most important phenomena in condensed matter physics: QHE, topological order, and superconductivity. To mention few concrete models, our result applies  to spin chains in the Haldane phase such as the AKLT model~\cite{Affleck:1988vr}, stabilizer codes~\cite{Bravyi:2011ea}, or the transverse field Ising model. We demonstrate the utility of the theorem by giving the first mathematical proof of Kubo's formula for interacting systems. 

Adiabatic theory was first applied to extended systems in connection to the QHE. In  a series of papers, Thouless et.~al.~\cite{TKNN, Thouless83, Thouless85} and Laughlin \cite{Laughlin} explained the quantization of Hall conductance by connecting the response coefficients to Berry's curvature via adiabatic response theory. The connection was further clarified in the works of Avron et.~al.~\cite{AvronSeilerSimon83, AvronSeiler85, ASS90}, and similar methods were employed to study the fractional quantum Hall effect \cite{Arovas}, topological insulators \cite{KaneMele}, and other novel quantum phases of matter \cite{Bradlyn16}. In all these works the validity of the adiabatic approximation for interacting systems is either taken for granted, or the scaling of the driving rate with the volume is unphysical.

In a separate line of research, the adiabatic approximation in many-body systems provides the basic theoretical framework for quantum annealing \cite{Finnila, Kadowaki, Farhi}. In this quantum simulation/computation method a ground state of an interacting Hamiltonian is reached from an initial separable ground state by slowly turning on the interactions. In a hard computational problem the gap of the Hamiltonian closes during the process and such crossing points give the dominant contribution to the error. If the gap does indeed not close, the target ground state can be efficiently simulated on a classical computer \cite{Osborne}. {Our result provides the quantum computational complexity of these algorithms, namely that target ground state is reached in a time of order $1$ in the number of spins.}

The classification of {gapped phases of matter is a question at the intersection of the above mentioned fields. Hastings and Wen \cite{HastingsWen, Wen} derived a basic property of such a gapped ground state phase: The local unitary equivalence of ground states along a path $H_s$, $0\leq s \leq 1$, of gapped Hamiltonians, which implies the smoothness of ground state expectation values.} We give a complementary dynamical result. By changing $H_0$ to $H_1$ in a time of order $\varepsilon^{-1}$, the expectation with respect to the solution of the driven Schr\"{o}dinger equation remains $\varepsilon$-close to the instantaneous ground state expectation, uniformly in the volume. Colloquially, the ground state of $H_0$ is transformed into the ground state of $H_1$ up to a small diabatic error.

\begin{figure}
\includegraphics[width=0.45\textwidth]{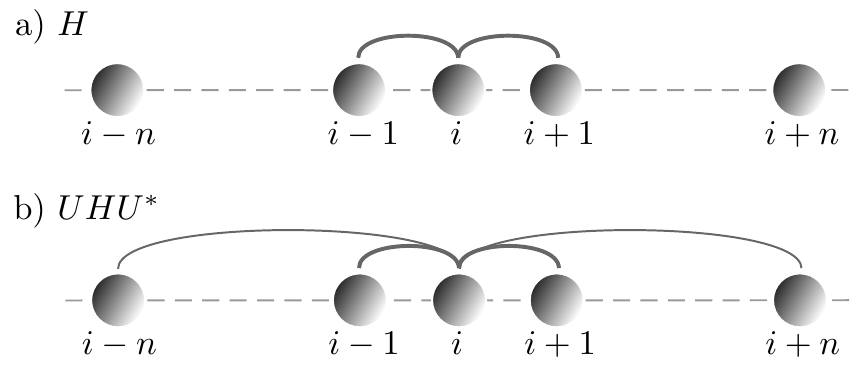}
\caption{\small The dressing transformation of order $n$. The Hamiltonian $H$
has nearest neighbors couplings (fig a.), and its ground state $\Omega$ is correlated on length scales of order $1$. The dressed Hamiltonian $UHU^*$ couples all sites in a ball of radius growing linearly with $n$ (fig b.). However the strength of the coupling between the sites $i$ and $i+n$ is of order $\varepsilon^n$. Correlations in the dressed ground state $\phi$ induced by $U$ are confined on length scales of order $n$.}
\label{fig}
\end{figure}

Quantum adiabatic theory has a long tradition going back to Born, Fock \cite{BornFock}, and Kato \cite{Kato50}. The latter realized that one can add a term $\varepsilon K_s$ to the Hamiltonian to make the adiabatic approximation exact. {Following~\cite{HastingsWen}, we shall call $K_s$ the generator of quasi-adiabatic (QA) evolution. The generator is not unique, and in a seminal paper \cite{HastingsWen} it was shown that there is an almost local choice of $K_s$, see also~\cite{Sven}}. In a time of order~$\varepsilon^{-1}$ such an added term {generically} contributes {an error of order $1$} compared to the real evolution.  We show that one can add an extensive quasi-local term $Y_{n,s}$ of order $\varepsilon^n$ to the Hamiltonian to get an evolution that follows a dressed ground state $\phi_{n,s}$ which is $\varepsilon$ close to the true ground state. In contrast to previous dressing constructions \cite{Berry90, Nenciu, Garrido, Hagedorn, Lidar} ours is local in both space and time, see Figure~\ref{fig}.
It is also specific to the many-body context that the solution of the Schr\"{o}dinger equation $\psi(s)$ is only $\varepsilon^{n-(d+1)}$ close to $\phi_{n,s}$, where $d$ is the dimension. In view of this, construction of local higher order correction terms is not used here to improve the adiabatic theorem but it is essential to obtain it.

In shortcuts to adiabaticity \cite{Berry09, Demirplak2005}, the correction term is employed as a control method. In this setting, $K_s$ is called a counterdiabatic driving and $Y_{n,s}$ is called a counterdiabatic driving of order $n$ \cite{Demirplak2008}. There is a great interest to engineer these correction terms and use them to reach a final ground state with higher fidelity. From this point of view we provide an explicit {local choice of $Y_{n,s}$}. Transitionless quantum driving for many-body quantum systems was studied in \cite{delCampo, Takahashi, Saberi}.

\subsection{Many-body adiabatic theorem}

We now explain the problem in details. Let $\Omega_s$ be the ground state of $H_s$. If we assume that
\begin{enumerate}
\item[(i)] the ground state energy is an isolated and simple eigenvalue separated from the rest of the spectra uniformly for all $0 \leq s  \leq 1$,
\item[(ii)] the ground state $\Omega_s$ is smooth enough,
 \end{enumerate}
then the adiabatic theorem \cite{Teufel, AE99, Jansen} says that the solution of the Schr\"{o}dinger equation in a rescaled time $s = t \varepsilon$, 
 \begin{equation}
 \label{eq:S}
 \varepsilon \frac{\d}{\d s}\psi(s)  = -iH_s \psi(s), \quad \psi(0) = \Omega_0,
 \end{equation}
 satisfies $||\psi(s) - e^{i \theta(\varepsilon,s)} \Omega_s|| \leq C \varepsilon$, where the constant $C$ depends on the family $H_s$, and $\theta(\varepsilon,s)$ contains a dynamical phase $\propto \varepsilon^{-1}$ and the more meaningful Berry phase.   
The rough estimate
$$
||\psi(s) - e^{i \theta(\varepsilon,s)}\Omega_s|| \approx\varepsilon \int_0^s \frac{||\dot{H}_{s'}||}{g_{s'}^2} \d s',
$$
where $g_s$ is the instantaneous  gap at time $s$, yields the volume dependence of $C$: Since the energy is extensive, the error grows as volume even though the gap remains open, marking the failure of the standard adiabatic estimates for extended systems.

 It is instructive to examine this in details for the non-interacting Hamiltonians $H_s=\sum_x D_{x,s}$, for example $D_{x,s}=\vec{h}_s \cdot \vec{\sigma}_x $ with some varying magnetic field. Then the solution $\psi(s)= \otimes_{x}\psi_x(s)$ of~(\ref{eq:S}) is a product state, and by the standard adiabatic theorem each $\psi_x(s)$ satisfies
$$
\psi_x(s) = e^{i\theta_x(\varepsilon,s)} \Omega_{x,s} + \mathcal{O}(\varepsilon),
$$
namely $\langle \psi_x(s) \str   e^{i\theta_x(\varepsilon,s)}\Omega_{x,s} \rangle = 1-\mathcal{O}(\varepsilon)  $. This however implies that the overlap of the tensor products decreases as $(1-\mathcal{O}(\varepsilon))^{V}$ with $V$ the volume. In the limit of an infinite volume, the vectors $\psi(s)$ and $\Omega_s$ become orthogonal, a phenomenon reminiscent of Anderson's orthogonality catastrophe~\cite{Lychkovskiy:2016aa}. This  shows that the breakdown of the standard adiabatic theorem cannot be argued away.
 
This consideration suggests the solution: The mismatch between $\psi(s)$ and $e^{i \theta(\varepsilon,s)}\Omega_s$ must be measured in a physically more meaningful way, typically by looking at expectation values of local operators. For independent spins the corresponding adiabatic theorem is immediate. 

The core of the problem is to deal with interactions. In the adiabatic setting, local interactions acting for a time of order $\varepsilon^{-1}$ yield a propagation over a distance of order $v_{LR} \varepsilon^{-1}$, where $v_{LR}$ is the Lieb-Robinson velocity~\cite{Lieb:1972ts, Nachtergaele:2006fd}. On the other hand to establish a local adiabatic theorem we need to show that all correlations beyond certain finite, $\varepsilon$-independent length scale, are negligible. 

Let us now give the precise phrasing.  For concreteness we consider a quantum spin system \cite{Mahan:2011aa,Bratteli:1997aa, Simon:1993aa} on the $d$-dimensional torus $\Lambda_L \equiv \bbZ^d/(L\bbZ)^d$ with a volume (number of lattice sites) $L^d$. The Hamiltonians are of the form
\begin{equation}
\label{eq:H}
H_s = \sum_{X \subset \Lambda_L} H_{X,s}.
\end{equation}
where the interaction term $H_{X,s}$ acts on the spins located in the subset $X$ only and it is smooth in $s$. We moreover assume that $H_s$ have finite range, i.e.\ $H_{X,s}=0$ whenever $\str X \str >r $, where the range $r$ is independent of volume. We say that $H_s$ is a gapped family if the spectral gap above the ground state energy remains open uniformly in $s$ and in the volume.

In order to measure the local size of an extensive operator $B=\sum_X B_X$, we shall use the norm
$$
\norm B\norm_{\text{loc}}:= \sup_x \sum_{X:X \ni x}\norm B_X\norm.
$$
Clearly, $\norm B\norm_{\text{loc}}$ may remain bounded when $L\to\infty$, in which case we call $B$ \emph{quasilocal}, even though the more precise statement is that `$B$ is \emph{a sum of} quasilocal terms'. We assume that $H_s$ is locally smooth, in the sense that 
$\norm \partial^k_s H_s \norm_{\mathrm{loc}}  \leq C_k$ for all $k$, and that the driving starts smoothly, $\partial^k_s H_s=0$ at $s=0$. 

Under these conditions we show that the dressed ground state $\phi_{n,s}$ of order $n$, constructed recursively below, follows the evolution up to order $n-d-1$.
 \begin{theorem}
 \label{thm:main}
 Let $H_{s}$ be a family of gapped, locally smooth, finite-range Hamiltonians of the form~(\ref{eq:H}) for $0 \leq s \leq 1$. 
  Then the solution $\psi(s)$ of the Schr\"{o}dinger equation (\ref{eq:S}) satisfies 
 \begin{equation}
 \label{maind}
 \vert \langle \psi(s) \str O  \psi(s) \rangle - \langle \phi_{n,s} \str O  \phi_{n,s} \rangle \vert \leq  C \varepsilon^{n-d-1}
 \end{equation}
and in particular 
  \begin{equation}
 \label{main}
 \vert \langle \psi(s) \str O  \psi(s) \rangle - \langle \Omega_s \str O  \Omega_s \rangle \vert \leq  C \varepsilon
 \end{equation}
  for any local observable $O$, and with constant $C$ independent of $s,L$.
 \end{theorem}

 To avoid discussing the dynamical phase, we henceforth assume $H_s \Omega_s = 0$. This can be always achieved by a time-dependent gauge transformation. We also incorporate the Berry phase into $\Omega_s$ by choosing $\langle{\Omega_s}| \partial_s \Omega_s \rangle = 0$.

Let us now develop the proof, using the intuition outlined above. We shall construct a quasilocal $A_s$ for which $\norm A_s \norm_{\text{loc}} = \mathcal{O}(\varepsilon)$ uniformly in $L$ and $s$, and define a unitary dressing transformation $U_s=\e^{-i A_s}$. Then, the vector $\phi_s=U_s\Omega_s$ is close to $\Omega_s$ in the sense that local expectation values will differ by an order $\varepsilon$ \footnote{Of course, the overlap of $\phi_s$ and $\Omega_s$ tends to zero with increasing volume}. Furthermore, $A_s$ will be chosen so that $\phi_s$ is the solution of a Schr\"{o}dinger equation with an additional driving term $Y_{s}$ that is quasilocal and $\varepsilon^n$-small:
 \begin{equation}
 \label{eq:Sdrive}
  \varepsilon  \frac{\d}{\d s}\phi_s  = - i (H_s+ Y_{s}) \phi_s, \quad \phi_0 = \Omega_0.
 \end{equation}
 The largest order $n$ that can be obtained is only bounded by the smoothness of the Hamiltonian. The vanishing of initial derivatives of the Hamiltonian ensures that $\phi_0 = \Omega_0$. Without this condition $\phi_0$ is only $\varepsilon$-close to $\Omega_0$ and such error cannot be controlled. We will prove that \eqref{eq:Sdrive} suffices to deduce \eqref{maind} but the idea is clear: the solution of the equation without counterdiabatic driving~(\ref{eq:S}) is close to the solution with driving. Although the ansatz of a dressing transformation echoes the standard constructions such as~\cite{Hagedorn}, its implementation is conceptually and technically of a very different nature, ensuring locality in \emph{both} time and space.

\subsubsection{The counter-diabatic driving $Y_n$}
Let us now explain how to derive \eqref{eq:Sdrive}. To avoid clutter, we drop the parameter $s$ since all quantities carry $s$-dependence. For $n=1$, $Y_1 = \varepsilon K$ does the job. Indeed, by the Lieb-Robinson bound, the QA generator is quasi-local and satisfies $\partial \Omega=-i K \Omega$, hence also 
$$
 \varepsilon \partial \Omega  = -i(H+\varepsilon K) \Omega
$$
which is indeed of the form \eqref{eq:Sdrive} but with $U=1$, i.e.\  $A=0$.  

We move now to general $n$ with $U=\e^{-i A}$ and the Ansatz $A=\sum_{j=1}^n\varepsilon^j A_j$. The following identity is the starting point of our analysis:
\begin{equation}
\label{eq: starting identity}
\varepsilon\partial\phi= -i[ \varepsilon i \partial U  U^\dagger  +  \varepsilon UKU^\dagger  +  (UHU^\dagger-H)  + H] \phi
\end{equation}
This follows simply by first noting that $\partial\phi= \partial U  U^\dagger \phi+ U\partial \Omega$, then using again the QA flow to resolve $\partial \Omega$ and adding $UHU^\dagger \phi= UH\Omega=0$. 
The first three of four terms in the bracket give the driving term $Y_n$.

\paragraph{The case $n=2$.}
We first check the claim for $n=2$, which will elucidate the main mechanism at work. 
Expanding the exponential in $U$, we find that $Y_2 \phi= \mathcal{O}_{\text{loc}}(\varepsilon^2) \phi +\varepsilon U(-i[A_1,H]\Omega +K\Omega) $ and hence $Y_2$ is of order $\varepsilon^2$ provided that  
\begin{equation} \label{eq: vanish first order}
-i[A_1,H]\Omega +K\Omega=0.
\end{equation}
Hence we need to choose $A_1$ so that
\begin{equation} \label{eq: inverse}
 A_1\Omega= iH^{-1}K\Omega + C_1 \Omega
\end{equation}
where $H^{-1}K\Omega$ is well-defined because $H$ has a gap and $\langle \Omega \str K \Omega\rangle=i\langle{\Omega}| \partial \Omega \rangle=0 $. The freedom to pick any constant $C_1$ will be crucial later on.  The key point is to solve \eqref{eq: inverse} with a quasi-local $A_1$. This can be achieved by using a trick that lies at the heart of the QA flow, namely the representation
\begin{equation} \label{eq: trick}
i H^{-1} K\Omega = \int \d t h(t) \e^{-i t H} K\Omega = \int \d t h(t)\tau_t(K) \Omega
\end{equation}
where $\tau_t(K)=\e^{-i t H} K \e^{i t H} $ and
 $h$ is a fast decreasing function with Fourier transform $\hat h(\omega)=\mathrm{i}(1/\omega+v(\omega))$ where $v$ is a real-valued function supported in $(0,g)$. Crucially, the right hand side of~\eqref{eq: trick} is of the form $L\Omega$ with a quasilocal $L$. Indeed, $h(t)$ is cut-off at large times while the Lieb-Robinson bound allows to localize $\tau_t(K)$. For proofs and examples of such $h$ we refer to \citep{Sven, HastingsWen}. In summary, we reach our goal by choosing 
 $$A_1=\int \d t h(t) \tau_t(K)+C_1.$$

\paragraph{General $n$.} For $n>2$, $Y_n\phi$ is of order $\mathcal{O}_{\text{loc}}(\varepsilon^n) \phi $, provided the $A_j$s satisfy higher-order analogues of \eqref{eq: vanish first order},
\begin{equation} \label{eq: general inverse vanish}
(M_p+N_p )\Omega=0
\end{equation} 
for $0\leq p\leq n-1$. Writing $\ad_A(B)\equiv [A,B]$,
\begin{align}
M_p&= \sum_{\mathbf{j}:d( \mathbf{j})=p} \frac{(-i)^k}{k!}\ad_{A_{j_k}}\ldots \ad_{A_{j_1}}(H) \label{eq: mp} \\
N_p&= i\sum_{\mathbf{j}:d( \mathbf{j})=p-1} \frac{i^k}{k!}\ad_{A_{j_k}}\ldots \ad_{A_{j_2}}(\partial A_{j_1})\label{eq: np}
\end{align}
for $p\geq 2$, where the sum is over finite index sequences $\mathbf{j}= (j_1,\ldots,j_k)$ and $d( \mathbf{j})= j_1+\ldots+j_k$. 
Provided that we have solved this problem already up to $p-1$, i.e.\ all $A_j, j\leq p-1$ can be chosen quasilocal, we can solve it for $A_p$. Indeed, 
the only term involving $A_p$ is the term corresponding to $\mathbf{j}=(p)$ in $M_p$ so we can recast \eqref{eq: general inverse vanish} as
\begin{equation} \nonumber
i H A_{p}\Omega + (L_p+ \partial A_{p-1}) \Omega=0
\end{equation}
where we used again $H\Omega=0$, and we separated the term $\partial A_{p-1}$, corresponding to $\mathbf{j}=(p-1)$ in $N_p$ to prepare for the sequel. Since $L_p$ consists entirely of iterated commutators of quasilocal terms, it is itself quasilocal. Analogously to the case $n=2$, the equation can be solved provided that 
\begin{equation} \label{eq: condition p}
\langle \Omega\str (L_p+ \partial A_{p-1}) \Omega \rangle=0,
\end{equation}
in which case $A_p$ is chosen as
\begin{equation} \label{eq: final expression ap}
 A_p=\int \d t h(t) \tau_t(L_p+\partial A_{p-1} )+C_p.
\end{equation}
To satisfy \eqref{eq: condition p}, we use the remaining freedom of choosing the constants $C_{j}$. More precisely, \eqref{eq: condition p} holds by taking it to be a solution of $\partial C_{p-1}=\langle \Omega\str (i[K,A_{p-1}] - L_p) \Omega \rangle$. 

To conclude, let us give an explicit expression for the remaining external driving term $Y_n$ for $n>1$. It is given by   $Y_n= U \tilde{Y}_n U^\dagger$ with
\begin{align}
 \widetilde{Y}_n = &\sum^{(n)}_{\mathbf{j}:d(\mathbf{j})\geq n } \varepsilon^{d( \mathbf{j})}\frac{(-i)^k}{k!}\ad_{A_{j_k}}\ldots \ad_{A_{j_1}}(H ) \nonumber \\ 
 & + i  \sum^{(n)}_{\mathbf{j}:d(\mathbf{j})\geq n } \varepsilon^{1+d( \mathbf{j})}\frac{i^k}{k!}\ad_{A_{j_k}}\ldots \ad_{A_{j_2}}(\partial A_{j_1}) 
\end{align}
where the superscript $(n)$ on the sums indicates that sequences $\mathbf{j}$ should be made with  $j_i<n$ only. This shows that $Y_n$ is indeed quasilocal, with $\norm Y_n \norm_{\text{loc}} = \mathcal{O}(\varepsilon^n)$. 

\subsubsection{Proof of the theorem}
Having constructed the counter-diabatic quasi-local terms $Y_{n,s}$, we proceed by employing  ordinary perturbation theory. Let $O(s,s')$ be the Heisenberg evolution from time $s'$ to $s$ corresponding to~(\ref{eq:S}) of a local observable  $O$. Then, by Duhamel's principle 
\begin{multline*}
\langle \psi(s)\str O \str \psi(s)\rangle = \langle\phi_s\str O \str \phi_s\rangle \\
 + \frac{i}{\varepsilon} \int_0^s \langle\phi_{s'}\str[{Y}_{n,s'}, O(s,s')] \phi_{s'}\rangle \d s'.
\end{multline*}
By the Lieb-Robinson bound $O(s,s')$ is supported in a ball of radius of order $\varepsilon^{-1}(s-s')$ up to an exponentially small tail \footnote{For any finite range observable $X$, $||[X,O(s,s')]|| \leq C \varepsilon^{-d}$, with $C$ dependent on $v_{LR}$ and $X,O$.}. Hence we get 
$
|\langle\phi_{s'}\str[{Y}_{n,s'}, O(s,s')] \phi_{s'}\rangle | \leq C \varepsilon^{n-d}.
$
This establishes \eqref{maind}
which in turn implies \eqref{main} using  $\langle\phi_s\str O \phi_s\rangle- \langle\Omega_s\str O \Omega_s\rangle=\mathcal{O}(\varepsilon)$.

\subsection{Linear Response Theory}

A major theoretical application of the expansion presented above is a rigorous proof of linear response theory. The response per unit volume of an  extensive observable $X$ is  to leading order proportional to a drive {$\varphi$},  $  L^{-d}\langle  \delta X \rangle  = f \delta \varphi + o(\delta\varphi)$,   with a response coefficient $f$ given by
\begin{equation}
\label{eq:f}
f = \lim_{\varepsilon \to 0} \lim_{L \to \infty} L^{-d} \frac{\langle\psi(s)\str X \psi(s)\rangle - \langle\Omega_s\str X \Omega_s\rangle}{\varepsilon \partial_s \varphi }.
\end{equation}
Here $\psi(s)$ is the solution of (\ref{eq:S}) with $H_s \equiv H_{\varphi_s}$. Though a canonical expression for $f$ is a textbook material \cite{Kubo:1957cl}, a mathematical proof of the existence of the limit is a notoriously hard problem \cite{Simon:1984aa}. Our results allows however to control this limit through the estimate 
$$
|\langle\psi\str X  \psi\rangle - \langle\Omega\str X  \Omega\rangle - i \varepsilon \langle\Omega\str [A_1,X] \Omega\rangle | \leq C L^d \varepsilon^2,
$$
with a volume independent constant $C$. From (\ref{eq:f}) we then have the Kubo linear response formula in a form
$$
f = i \langle [A_1,X]\rangle
$$
where $\langle \cdot \rangle$ denotes the ground state expectation value per unit volume. One then recovers various standard response expressions \cite{AvronSeilerSimon83, BerryRobbins, Gritsev, AFG, Bradlyn} by plugging an appropriate choice of $X$ and $H_s$.
This is the first mathematical proof of linear response theory for a broad class of  interacting systems in the correct order of limits, i.e.~the thermodynamic limit first. See \cite{Jaksic:2006fv, Jaksic:2010aa, AbouSalem:2005kr, klein2007mott,Bru:2016hv,Bru:2017aa} for other advances on this problem.

This result complements first principle proofs of the quantization of Hall conductance in interacting systems \cite{Giuliani:2016gn, HastingsMichalakis}.

\subsection{Nonperturbative effects}

As indicated in \eqref{maind}, the diabatic error can be made non-perturbative (smaller than any order in $\varepsilon$) provided $H_s$ becomes constant at the end of the protocol, in which case $\phi_{s} = \Omega_s$ at $s=1$. In the standard, {one-body} adiabatic theorem, this remaining error is of order $\e^{- \min_s g_{s}/\varepsilon}$. 
We show that in the many-body case the scaling depends on the dimension.
Due to the time-evolution in \eqref{eq: final expression ap}, the typical range of the local terms in $A_p$ cannot be smaller than $((v_{LR}/g) p)^d$. If we assume that the term $(1/2)[A_{p-1},[A_1,H]]$ in $L_p$ gives the leading contribution to the right-hand-side of \eqref{eq: final expression ap}, then we find \footnote{If $A=\sum A_X, B=\sum B_X$ consist of local terms whose { range} is $r_A,r_B$ respectively (i.e.,  $A_X=0$ unless $\str X\str \leq r_A$) then $\norm [A,B] \norm_{\text{loc}} \leq  2\max\{r_A,r_B\}\norm A \norm_{\text{loc}} \norm B\norm_{\text{loc}} $} that
$$
\norm A_p \norm_{\text{loc}} \sim p^d   \norm H \norm_{\text{loc}}  \norm A_1 \norm_{\text{loc}}  \norm A_{p-1} \norm_{\text{loc}}
$$
hence $\norm A_p \norm_{\text{loc}} \sim (p !)^d$. This suggests that the optimal order $n$ at which to stop the adiabatic expansion, found by $\varepsilon^n \norm A_n \norm_{\text{loc}} =1$, is given by $n= c/\sqrt[d]{\varepsilon}$ for some constant $c$, and the resulting error term is then $\e^{-c/\sqrt[d]{\varepsilon}}$. Although this estimate may underestimate the diabatic term in general (it does so in $d=0$), it does strongly suggest that in $d>1$, the one-body estimate is wrong. This dimension dependent scaling is the most striking manifestation of the novel dressing construction that we describe in this letter. The scaling can be probed in cold atom experiments \cite{Esslinger}.

An interesting related point that is absent in the single-body framework is that the diabatic term contains long-range contributions, originating from entangled excitations created during the driving. These excitations can propagate for time of order~$\varepsilon^{-1}$ and hence introduce a correlation over lengths of at most~$\varepsilon^{-1}$ in the dressed ground state $\phi$, to be contrasted with the  static correlation length {of order~$g^{-1}$} in the gapped $\Omega$, see Fig.~\ref{fig}. 

Let us illustrate this divergent correlation length in the case of the transverse-field Ising chain, $H_s = -\sum_{i=1}^L\left(h_s \sigma_i^z + \sigma_i^x\sigma_{i+1}^x\right)$. We choose $h\to+\infty$ for large $\vert s \vert$, while $h$ has a minimum $h_0>1$. The system is initially in the $h=\infty $ ground state ${\footnotesize \vert \Omega_{-\infty} \rangle =\vert \uparrow \ldots \uparrow\rangle}$, and so is its final state to all orders, with corrections beyond perturbation theory. When $h\sim h_0$ each mode undergoes an avoided crossing with a gap given by $\gamma_k^2 = 4(h_0^2+1-2h_0\cos(k))$ yielding a tunnelling probability $p_k$ given by the Landau-Zener formula. The density of excitations above the final ground state in the $L\to\infty$-limit is thus given \cite{Dziarmaga:2010da} by
\begin{equation}\label{density}
\rho(\varepsilon) =  L^{-1}\sum_{k} p_k \approx \frac{\sqrt{\varepsilon/h_0}}{4\sqrt 2 \pi}e^{-\frac{8\pi}{\varepsilon}(h_0-1)^2}.
\end{equation}
The correlations arising from the created entangled pairs of excitations can be read from $\langle \sigma_i^z\sigma_{i+l}^z\rangle - \langle \sigma_i^z\rangle\langle \sigma_{i+l}^z\rangle$ at time $t=\infty$, whose leading contribution is
\begin{equation*}
L^{-2}\sum_{k,q} p_k p_q e^{i l (p-k)}\approx C \rho(\varepsilon) e^{-\frac{l^2\varepsilon }{16h_0}}.
\end{equation*}
As discussed above, this corresponds to a correlation length of order $\varepsilon^{-1/2}$ corresponding to dispersive free exciton pairs over a time of order $\varepsilon^{-1}$.

\subsection{Conclusion}

We prove an adiabatic theorem appropriate for extended quantum spin systems. The main idea is to introduce a dressing operator $U$ to describe the adiabatic cloud and to look for a local generator $A$ for $U$ in the form of an asymptotic expansion in powers of the adiabaticity parameter $\varepsilon$. This yields also a local counterdiabatic driving term. We show that the density of non-perturbative errors have a non-standard dimension dependent behavior. In an exactly solvable model we compute  the spatial correlations of these errors. We expect that the perturbation expansion will find further applications in mathematical/theoretical physics beyond the proof of Kubo's formula that we presented here, for example in the recent exciting developments in the adiabatic theory for periodically driven systems \cite{Abanin, Polkovnikov}.

\begin{acknowledgments}
\subsection{Acknowledgments}
W.D.R. and M.F. benefit from funding by the InterUniversity Attraction Pole DYGEST (Belspo, Phase VII/18). W.D.R also acknowledges financial  support from the  DFG
(German Research Fund) and the FWO (Flemish Research Fund).
\end{acknowledgments}

\end{document}